\documentclass[10pt,letterpaper]{article}
\usepackage{opex3}
\usepackage{cite}
\usepackage{graphicx}

\begin{document}

\title{Thermo-optic and thermal expansion coefficients of RTP and KTP crystals over 300-350 K}

\author{Arlee V. Smith$^{1,*}$, Jesse J. Smith$^1$, and Binh T. Do$^2$}

\address{$^1$AS-Photonics, 6916 Montgomery Blvd. NE, Ste. B8, Albuquerque, NM 87109, USA\\
$^2$Ball Aerospace Technologies, 2201 Buena Vista Dr. SE, Ste. 100, Albuquerque, NM 87106, USA\\
$^*$arlee.smith@as-photonics.com}

\begin{abstract*}
{We report new measurements of the thermal expansion and thermo-optic coefficients of RbTiOPO$_4$ (RTP)  and KTiOPO$_4$ (KTP) crystals over the temperature range 300-350 K. For RTP/KTP our coefficients of linear thermal expansion at 305 K are: $\alpha_x=9.44/7.88\times 10^{-6}$/K, $\alpha_y=12.49/9.48\times 10^{-6}$/K, $\alpha_z=-4.16/0.02\times 10^{-6}$/K. Our normalized thermo-optic coefficients $\beta=(1/n)dn/dT$ at 632.8 nm and 305 K are: $\beta_x=5.39/3.78\times 10^{-6}$/K, $\beta_y=7.11/5.24\times 10^{-6}$/K, $\beta_z=12.35/9.34\times 10^{-6}$/K.}
\end{abstract*}


\section{Introduction}

Thermal effects often dictate the performance limits of lasers and nonlinear optical devices. Two effects are particularly important, the thermo-optic effect or change in refractive index with temperature, and thermal expansion. They both affect thermal lensing, the thermo-optic effect by forming a gradient index lens, and thermal expansion by creating bulges on the ends of the crystal and by strain-induced changes in the refractive index. The thermo-optic effect also determines the rate of temperature tuning for angle phase matched nonlinear crystals, while both the thermo-optic effect and thermal expansion determine the temperature tuning of quasi phase matched crystals. Accurate and precise values of both coefficients are necessary for general calculations of temperature tuning. Significant disagreement among previously reported values of the thermo-optic and thermal expansion coefficients of RTP (RbTiOPO$_4$) and KTP (KTiOPO$_4$) make such calculations uncertain. Here we report new measurements of the thermal expansion coefficient $\alpha$ and the thermo-optic coefficient $\beta$ at 632.8 nm and over the temperature range $(300 K<T<350 K)$. Our $\beta$ values at only one wavelength are insufficient for phase matching calculations, but they provide a strong check on previous measurements. 

\section{Methods}

In anisotropic crystals the thermal expansion and index of refractive are direction dependent. Each is described by an associated $3\times 3$ tensor. In their respective principal frames, each tensor is diagonal and the three diagonal elements are the principal values of the linear thermal expansion coefficients $\alpha$ and the normalized thermo-optic coefficients $\beta$. The crystals RTP and KTP have orthorhombic symmetry, so the principal frames for $\alpha$ and $\beta$ are co-aligned with the three crystalline axes\cite{Nye,Smithbook}. There are three different principal expansion coefficients and three different principal thermo-optic coefficients. 

Our measurements are made for light propagating along one of the principal axes and polarized parallel to another principal axis. We use the subscript $i$ to indicate the propagation direction and $j$ to indicate the polarization direction. For thermo-optic measurements we measure the interference of reflections from the input and output faces of crystal. For the thermal expansion measurements we use the one-pillar or the two-pillar method in which one or two crystal samples are used as spacers between silica optical wedges and the interference pattern is formed by light reflected from the two faces of the wedges that are in contact with the crystal faces. The same temperature controlled oven is used for both measurements. The setups used for the two measurements are diagrammed in Fig. 1.

The thermal expansion coefficient $\alpha$ and the thermo-optic coefficient $\beta$ both are zero at 0 K and gradually rise with temperature\cite{Callen,Ter-Gabrielyan}. This suggests that over the 300-350 K range covered in our measurements both $\alpha$ and $\beta$ can change noticeably with temperature. We approximate this temperature dependence using two-term expansions. Defining $\Delta T$ by ($\Delta T = T - T_\circ$), where $T_\circ$ is a reference temperature (which we take to be 305 K), we express the temperature-dependent length and refractive index in the form
\begin{equation}\label{eqn:length_expansion}
L_i(T) = L_{\circ i} (1 + a_{\alpha i} \Delta T + \frac{1}{2} b_{\alpha i} \Delta T^2 + ...)
\end{equation}
\begin{equation}\label{eqn:n_expansion}
n_j(T) = n_{\circ j} (1 + a_{nj} \Delta T + \frac{1}{2}b_{nj} \Delta T^2 + ...)\label{eqn:n_air_expansion1},
\end{equation}
where $L_{\circ i}$ refers to the crystal length along the propagation axis $i$ at $T_\circ$, and $n_{\circ j}$ refers to the refractive index of $j$ polarized light at $T_\circ$. The definitions of $\alpha_i$ and $\beta_j$ are
\begin{equation}
\alpha_i = \frac{1}{L_{\circ i}} \frac{d L_i}{d T} =  a_{\alpha i} + b_{\alpha i} \Delta T+...,
\end{equation}
\begin{equation}
\beta_j = \frac{1}{n_{\circ j}} \frac{d n_j}{d T} = a_{nj} + b_{nj} \Delta T+....
\end{equation}
The round-trip phase for two passes through the crystal $\phi_{ij} (T)$ can be expressed in the form
\begin{eqnarray}
\phi_{ij}(T) &=& 2 k_\circ n_j(T) L_i(T) \\\nonumber
&=& \phi_{\circ ij} \bigg(1 + a_{\alpha i} \Delta T + \frac{1}{2} b_{\alpha i} \Delta T^2+..\bigg) \bigg(1 + a_{nj} \Delta T + \frac{1}{2} b_{nj} \Delta T^2+..\bigg) \\\nonumber
&=& \phi_{\circ ij} \bigg[ 1 + (a_{nj} + a_{\alpha i})\Delta T + \bigg(\frac{1}{2}b_{nj} + \frac{1}{2}b_{\alpha i} + a_{\alpha i} a_{nj} \bigg)\Delta T^2 + ... \bigg]\nonumber
\end{eqnarray}
where $k_\circ = \omega / c$ and $\phi_{\circ ij} = 2 k_\circ n_{\circ j}L_{\circ i}$ is the round-trip phase at $T=T_\circ$. We will see that the cross-term $(a_{\alpha i} a_{nj})$ is usually negligible compared to $(b_{nj}+b_{\alpha i})$, so we drop it, leaving
\begin{equation}\label{eq:phi_of_T}
\phi_{ij}(T) = \phi_{\circ ij} \bigg[ 1 + (a_{nj} + a_{\alpha i})\Delta T + \frac{1}{2} \bigg(b_{nj} + b_{\alpha i} \bigg)\Delta T^2 +... \bigg].
\end{equation}
We define $\gamma_{\,ij}$, the normalized optical path length rate of change with temperature, as
\begin{equation}
\gamma_{\,ij} = \frac{1}{\phi_{\circ ij}} \frac{\partial \phi_{ij}(T)}{\partial T} = (a_{\alpha i} + b_{\alpha i}\Delta T+..)+(a_{nj} + b_{nj} \Delta T+..),
\end{equation}
or
\begin{equation}
\gamma_{\,ij} = \alpha_i+\beta_j.
\end{equation}

To determine the values of $\alpha_i$ and $\beta_j$ we make two measurements. A measurement of $\gamma^{\;\rm crystal}_{ij}$ is performed using the arrangement in Fig.~\ref{fig:experimental_setup}(a) and a measurement of $\gamma^{\;\rm air}_i$ is performed using the arrangement in Fig.~\ref{fig:experimental_setup}(b). The crystal is placed in an oven and its temperature is gradually stepped up and down while the strength of the reflected light is monitored. Interference fringes are curve fit assuming $\alpha$ and $\beta$ have the form $(a+b\Delta T)$. Air is not birefringent so we drop the polarization subscript $j$ for the air values. The two measured $\gamma$ values are related to $\alpha$ and $\beta$ by
\begin{equation}\label{eqn.gamma_c} 
\gamma^{\;\rm crystal}_{\,ij} = \alpha^{\;\rm crystal}_i + \beta^{\;\rm crystal}_j,
\end{equation}
\begin{equation}\label{eqn.gamma_a} 
\gamma^{\;\rm air}_i = \alpha^{\;\rm crystal}_i + \beta^{\rm air}.
\end{equation}
The value of $\beta^{\rm air}$ can be computed for a standard atmospheric composition, taking into account the local barometric pressure and its variation during a measurement. From the measured $\gamma_i^{\;\rm air}$ and the computed $\beta^{\;\rm air}$ we find $\alpha_i^{\;\rm crystal}$ using Eq. (\ref{eqn.gamma_a}). Using this $\alpha_i^{\;crystal}$ and the measured value of $\gamma_{\,ij}^{\;\rm crystal}$ in Eq. (\ref{eqn.gamma_c}) yields $\beta_j^{\;\rm crystal}$. In the following we abbreviate $crystal$ and $air$ with the superscripts $c$ and $a$.

\section{Measurement methods}

\begin{figure}[htbp]
\centering
\includegraphics[width=\textwidth]{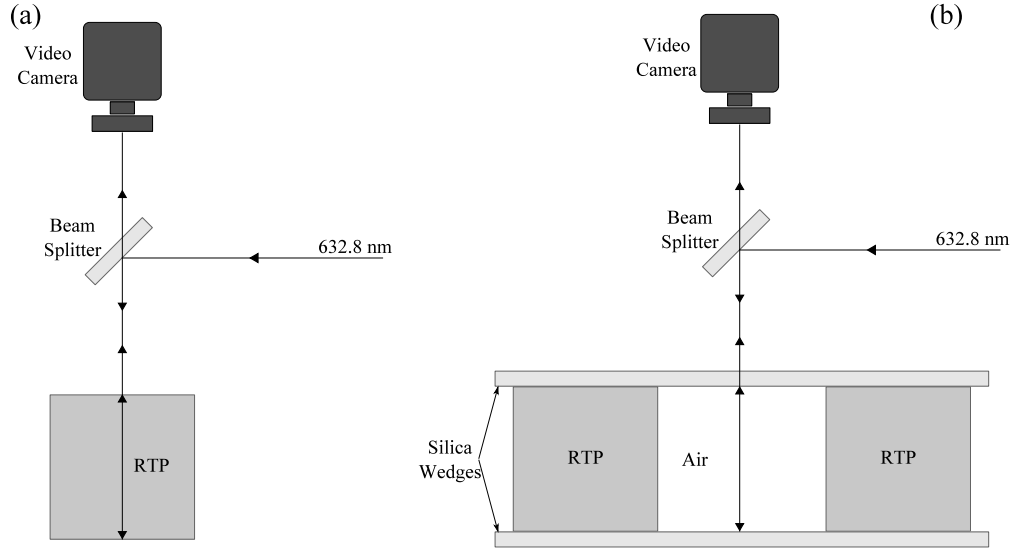}
\caption{\label{fig:experimental_setup} Diagram (a) shows a $\gamma_{\,ij}^{\;c}$ measurement. The light propagates along the $i$ axis and is $j$ polarized. Diagram (b) shows a $\gamma_i^{\;a}$ measurement. The light propagates parallel to the $i$ axis and interference occurs between reflections from the inner wedge surfaces. The crystals and silica wedges are uncoated. The RTP samples are 5.085 mm long, the KTP samples are 4.95 mm long, the silica wedges are 19 mm diameter, 6.35 mm thick and have a wedge of 30 arc-minutes.}
\end{figure}

We use six RTP samples from Cristal Laser SA, with pairs of crystals cut for propagation along each principal axis, each $5.08\pm 0.03$ mm long. The $\gamma_i^{\;a}$ measurements use paired crystals sandwiched between fused silica wedges as shown in diagram (b) of Fig. \ref{fig:experimental_setup}. We also have two KTP samples from Crystal Associates, both cut to a 4.88 mm length parallel to $z$. The dimensions in the $x$ and $y$ directions are 4.95 mm. All six sides of the KTP samples are polished. The $\gamma_i^{\;a}$ measurements for KTP use a single crystal in place of the crystal pair shown in Fig. 1b. The silica wedges are not optically contacted to the crystals in either the RTP or the KTP measurements.

Our copper oven is a thick cylindrical sleeve, 60 mm long, with a 22.2 mm inner diameter and a 44.4 mm outer diameter. A 6 mm thick sleeve of Teflon insulates the outer wall of the oven on the side and ends. An anti reflection coated silica window is placed on top of the oven to limit convective air currents. The oven is not sealed, so the air pressure inside the oven is in equilibrium with that in the laboratory. The crystal sits on a pedestal to position it close to the center of the oven.

A Cryocon Model 22 temperature controller monitors the oven temperature using a 100~$\Omega$ platinum resistive temperature sensor embedded in the oven wall. The temperature controller measures the drop in potential across the platinum resistor with a 1 mA excitation current in a four-wire measurement. The Cryocon Model 22 manual states the absolute temperature measurement accuracy is 6.2 mK at 300 K, with 4.7 mK resolution. The temperature controller is a proportional-integral-derivative (PID) type and varies the heater power to reach a temperature setpoint. Four cartridge heaters (each rated for 80 W at 120 V) are inserted in holes in the oven wall parallel to the bore of the oven. We calibrated a K-type thermocouple read by a Fluke 51 II thermometer by placing it next to the platinum resistor. The thermocouple was then moved to the location of the crystal to compare the crystal temperature with the set point. The difference was less than 0.8 K across the 300-350 K measurement range.

The 632.8 nm HeNe laser is polarized but not frequency stabilized. Its cavity is approximately 600 mm long so mode hops lead to a $\pm 0.75$\% shift of the interference fringes from the 5 mm long crystals. In most cases there are ten or so fringes over the full temperature scan so this shift contributes little to the measurement error. Fluctuations of a few percent in laser power, perhaps associated with mode hops, also add some uncertainty in the fringe fitting. We monitor the reflected laser light using an 8-bit black and white CCD camera with image resolution of 640$\times$480 at 30 frames per second. Its image shows the quality of the interference fringe and also provides a visual indication of accidental beam tilts. We illuminate a large area of its detector to insure the interference fringes caused by the glass window covering the CCD sensor are averaged. We sum the total signal from all of the CCD pixels to form the reflected light signal.

During all $\alpha$ measurements we continuously monitor the barometric pressure using a Freescale Semiconductor Xtrinsic MPL3115A2 I$^2$C Precision Altimeter which gives a 20-bit measurement of pressure in Pascals, with absolute accuracy of $\pm 0.4$ kPa and relative accuracy of $\pm 0.05$ kPa. A microcontroller logs the pressure once per minute.

We created a LabView virtual instrument to manage the temperature controller's setpoint and to collect images from the video camera. For each set point we wait several minutes for the temperature sensor to reach the set point and for the sample to thermally equilibrate to the oven. Typical temperature increments are 0.1-3 K, and the scan range is always 300-350 K. For each set point we record the temperature controller's reading of the temperature sensor, and take 10 images from the camera spread over 30 seconds.

\begin{figure}[htpb]
\centering
\includegraphics[width=0.95\textwidth]{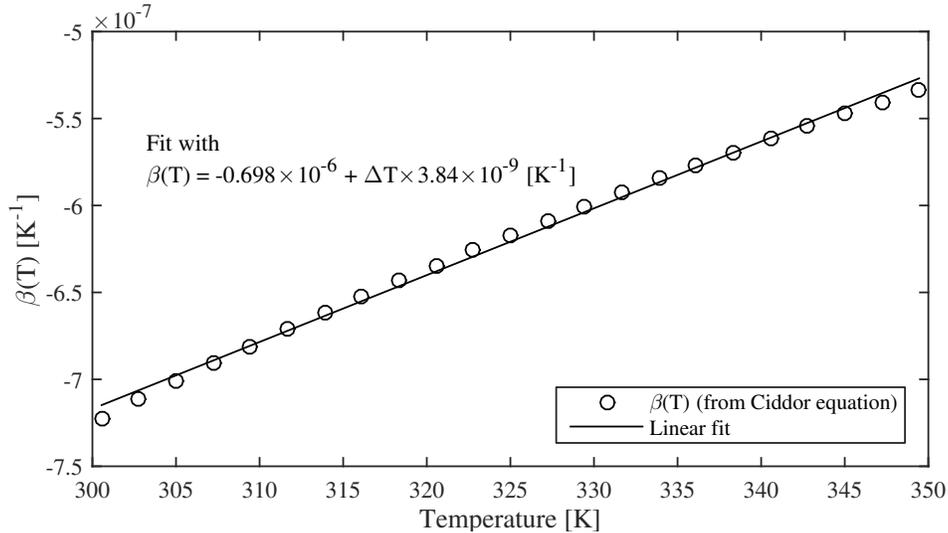}
\caption{\label{fig:air_beta}$\beta^{air}$ as a function of temperature, determined from the Ciddor equation\cite{Ciddor}. The linear fit assumes the form $\beta=a+b(T-305)$. Note the negative sign of the $\beta$ axis.}
\end{figure}

According to Eq. (\ref{eqn.gamma_a}) we must know $\beta^a$ to derive $\alpha_i^c$ from a measurement of $\gamma_i^{\;a}$. We calculated the refractive index of air over the range 300-350 K using the Ciddor equation\cite{Ciddor} described in the NIST Engineering Metrology Toolbox\cite{NIST}. Because our laboratory elevation is 1600 m, we defined our standard atmospheric pressure as $P_\circ = 83.0$ kPa. From the NIST calculator, the refractive index of air at 305 K, 632.8 nm, relative humidity of 25\% and $P_\circ$ is $n_{\circ a} = 1.0002135$. This means that at 305 K air adds approximately 21 radians of phase to a 10 mm round trip. The change in the refractive index of air with temperature over our measurement range is due largely to rarefaction. In applying the Ciddor equation we specify a constant water vapor partial pressure that corresponds to 45\% relative humidity at 298 K. Numerically differentiating the refractive index with respect to temperature, we find that the normalized thermo-optic change in refractive index is well fit by
\begin{equation}\label{eq.air_beta}
\beta^a = \frac{1}{n_\circ^a}\frac{d n^a}{d T} = -0.698 \times 10^{-6} + 3.84 \times 10^{-9} (T - 305 K).
\end{equation}
Figure \ref{fig:air_beta} compares the $\beta^a(T)$ calculated from the Ciddor equation and the linear fit of Eq.~(\ref{eq.air_beta}). We find the coefficients in Eq.~(\ref{eq.air_beta}) vary insignificantly over the relative humidity range 0-50\% near room temperature, and our measurements were performed at room temperature relative humidities of approximately 20\%.


To extract $\gamma_i^{\;a}$ from our data, we fit the reflected signal using a function of the form
\begin{equation}
\label{eqn:gamma_a_fit}f(T) = a_f \times \Big[ 0.5 + 0.5 \sin \big( b_f + c_f \Delta T + \frac{1}{2} d_f \Delta T^2 \big) \Big] + e_f,
\end{equation}
so the fit phase is
\begin{equation}
\phi_f = b_f + c_f \Delta T + \frac{1}{2} d_f \Delta T^2.
\end{equation} 
Differentiating $\phi_f$ with respect to temperature and dividing by $\phi_{\circ}^a = 2 k_\circ L_\circ n_{\circ}^a$ gives
\begin{equation}
\gamma_i^{\;a} = \frac{1}{\phi_{\circ}^a} \frac{\partial \phi_f}{\partial T} = \frac{c_f + d_f \Delta T}{\phi_{\circ}^a}=\alpha_i^{\;c} + \beta^a.
\end{equation}
Subtracting $\beta^a$ given by Eq. (\ref{eq.air_beta}) from $\gamma_i^{\;a}$ yields $\alpha_i^{\;c}$.
We use discrete temperature steps, but since we must make many measurements with many crystals in many orientations we don't take a uniform number of measurements per fringe period. In cases with short temperature periods (large $\gamma$ values), we might use only a few points per period $-$ although the number is always well above the Nyquist limit. 

Usually changes in atmospheric pressure during a run have negligible effect on our measurements, but occasionally the atmospheric pressure changes by as much as 2\%, corresponding to a round-trip phase change of 0.4 radians. We routinely precompensate our data using the measured barometric pressure by slightly adjusting the measured temperature by an amount corresponding to the pressure induced phase shift before computing the $\phi(T)$ fit. The temperature correction is
\begin{equation}
\label{eq:adjusting_for_pressure_variation}\delta T (P) = \pm \frac{P - P_\circ}{P_\circ}\ \frac{2 L_\circ k_\circ (n_{\circ}^a - 1)}{2 \pi}\ T_{\rm period},
\end{equation}
where $T_{\rm period}$ is the measured approximate period of the reflectivity pattern and the $\pm$ sign is the same sign as $\gamma_i^{\;a}$.

We find our $\gamma_{\,ij}^{\;c}$ measurements are highly reproducible, with run-to-run variations less than 0.05 in $a$ and less than 0.5/K in $b$. The $\gamma_i^{\;a}$ runs are less reproducible. To reduce the uncertainty we typically take several temperature scans and average the results. We think the run-to-run variations are due to small changes in the spacing between the crystal and the wedge(s). The wedge(s) are not optically contacted to the crystals so there is some position creep due to the difference in expansion rates of the silica wedges and the crystal. Contamination on the crystal or wedge surfaces may also contribute. We find no significant difference between the one-pillar and two-pillar measurements in this regard. Other potential sources of measurement uncertainty include slight displacements of the laser beam and slight tilts of the oven and crystal as they heat up, but these should be the similar in the much more reproducible $\gamma_{ij}^{\;c}$ runs, so we think they do not contribute to the variations.

\section{RTP results}

\begin{figure}[htbp]
\centering
\includegraphics[width=0.95\textwidth]{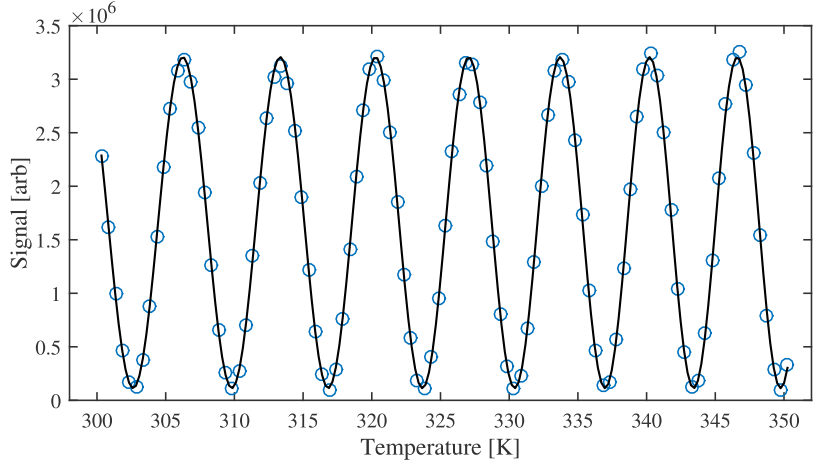}
\caption{\label{fig:gamma_xa_up1}Typical temperature scan for measurement of $\gamma_x^{\;a}$. The circles are the reflected signal; the solid curve is a best fit to the form of Eq.~\ref{eqn:gamma_a_fit} with five free parameters $a_f$ through $e_f$.}
\end{figure}

\begin{figure}[htpb]
\centering
\includegraphics[width=0.95\textwidth]{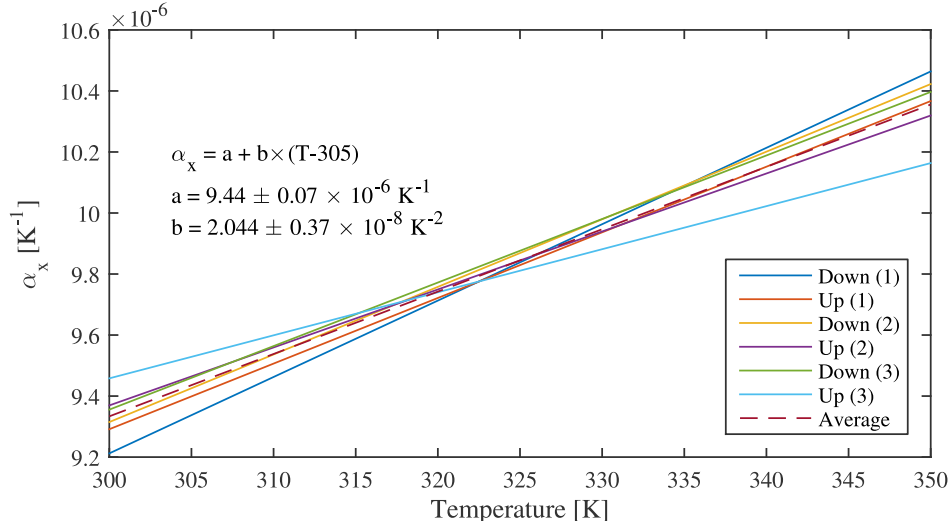}
\caption{\label{fig:rtp_alpha_x}$\alpha_x(T)$ as a function of temperature for RTP, determined from $\gamma_x^{\;a}$ using $\alpha_x = \gamma_x^{\;a} - \beta^a$. Each line represents a single temperature scan, and the uncertainties in $a$ and $b$ are determined from the statistical spread in $a$ and $b$. The dashed line indicates the average, given by the inset equation.}
\end{figure}

Figure~\ref{fig:gamma_xa_up1} shows a typical fringe measurement for a single temperature scan, in this case for $\gamma_{x}^{\;a}$. For a $\gamma_i^{\;a}$ measurement we run several temperature scans in both directions and derive statistical uncertainties of the $a$ and $b$ coefficients. Figure \ref{fig:rtp_alpha_x} shows an example of the derived $\alpha_x(T)$ values, including the $a$ and $b$ uncertainties.

Table \ref{tab:gamma_results} summarizes our nine independent $\gamma$ measurements for RTP. Our measurement method does not determine the sign of $\gamma$, so the signs of $\alpha_i$ are ambiguous. However, we know from measurements of the crystal lattice parameters as a function of temperature made by Delarue {\it et al}\cite{Delarue} that $\alpha_z$ is negative, and $\alpha_x$ and $\alpha_y$ are positive. The sign of $\gamma_{z}^{\;a}$ is also confirmed by the sign of the fringe shift observed for changing barometric pressure.

\begin{table}[htbp]
\centering
\caption{\label{tab:gamma_results} Best-fit values for $\gamma_{\,ij}$ and $\gamma_i^{\;a}$ for RTP at 632.8 nm over 300-350 K. Here $i$ refers to the propagation direction and $j$ refers to the polarization direction. We assume $\gamma(T) = a + b\times (T - 305 K)$. $\beta^a$ is the calculated thermo-optic coefficient for air at our laboratory elevation of 1600 m. We use $[n_x,n_y,n_z]=[1.7902,1.8014,1.8889]$ in deriving $\gamma_{\,ij}$.}
\begin{tabular}{c c c c}\hline
Parameter & a [K$^{-1}$] & b [K$^{-2}$] & $\frac{a}{b}$[K]\\ \hline
$\gamma_{\,xy}$ & $16.64 \times 10^{-6}$ & $44.92 \times 10^{-9}$ &370\\
$\gamma_{\,yz}$ & $24.78 \times 10^{-6}$ & $76.36 \times 10^{-9}$ &325\\
$\gamma_{\,zx}$ & $1.161 \times 10^{-6}$ & $9.364 \times 10^{-9}$ &124\\
$\gamma_{\,zy}$ & $2.862 \times 10^{-6}$ & $19.01 \times 10^{-9}$ &151\\
$\gamma_{\,yx}$ & $17.96 \times 10^{-6}$ & $40.81 \times 10^{-9}$ &440\\
$\gamma_{\,xz}$ & $21.85 \times 10^{-6}$ & $72.33 \times 10^{-9}$ &302\\
$\gamma_x^{\;a}$ & $8.737 \times 10^{-6}$ & $24.28 \times 10^{-9}$  &360\\
$\gamma_y^{\;a}$ & $11.80 \times 10^{-6}$ & $23.86 \times 10^{-9}$  &495\\
$\gamma_z^{\;a}$ & $-4.861 \times 10^{-6}$ & $-8.689 \times 10^{-9}$&559\\
$\beta^a   $  & $-0.699 \times 10^{-6}$ & $ 3.83 \times 10^{-9}$&\\
\hline
\end{tabular}
\end{table}

\begin{figure}[htbp]
\centering
\includegraphics[width=0.85\textwidth]{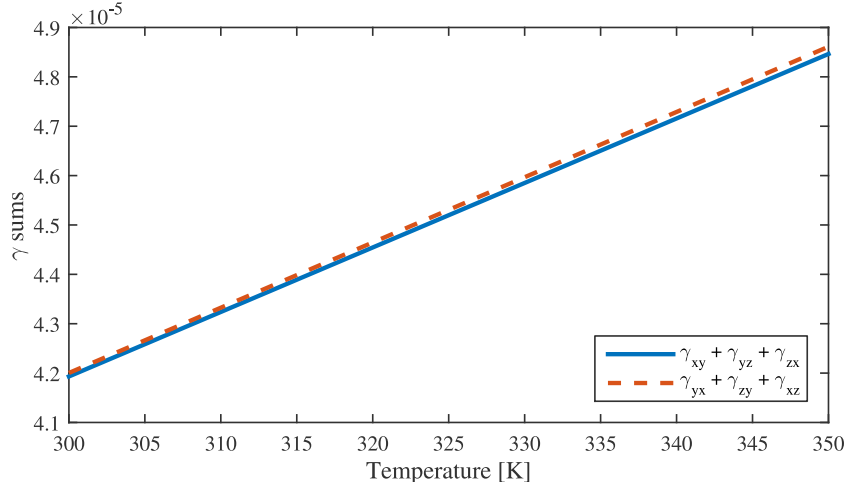}
\caption{\label{fig:gamma_check}Plot of the two $\gamma$ sums indicated by the caption, where the RTP $\gamma$s are taken from Table \ref{tab:gamma_results}. For perfect measurements, the two sums should be equal over the entire temperature range. Our measured values differ by 0.18-0.3\% over the 50 degree temperature range.}
\end{figure}

We measure all six $\gamma_{\,ij}$ values which is redundant since
\begin{equation}\label{eq.redundancy}
\gamma_{\,xy} + \gamma_{\,yz} + \gamma_{\,zx} = \gamma_{\,yx} + \gamma_{\,zy} + \gamma_{\,xz}.
\end{equation}
The sums on both sides of this equation are equal to ($\alpha_x+\alpha_y+\alpha_z+\beta_x+\beta_y+\beta_z$). The equality of the two sums in Eq. (\ref{eq.redundancy}) provides a check on the quality of our measured $\gamma_{\,ij}$ values. In Figure \ref{fig:gamma_check} we compare the two sums, showing that they are equal to within 1\%. The statistical variation of the individual $\gamma_{\,ij}$ measurements is less than 1\%.

In our tables of $\gamma$, $\alpha$, and $\beta$ we include the ratio $(a/b)$. If these quantities were zero at 0 K and grew linearly we would expect this ratio to be approximately 300 K at 300 K. Linear growth is not realistic, but nevertheless the value of $(a/b)$ offers a sensibility check worth noting.

To find $\alpha_i$, we subtract $\beta^a$ from $\gamma_{i}^{\;a}$ in Table \ref{tab:gamma_results}. These $\alpha_i$ values are summarized in Table \ref{tab:alpha_results_RTP}. 
\begin{table}[htbp]
\centering
\caption{\label{tab:alpha_results_RTP}$\alpha_i$ values for RTP in the form $\alpha = a + b \times (T - 305)$. These $\alpha_i$ are computed from the measured $\gamma_i^{\;a}$ using $\alpha_i = \gamma_{i}^{\;a} - \beta^a$. The uncertainties reflect the run-to-run variations in $\gamma_i^{\;a}$.}
\begin{tabular}{c c c c }\hline
Parameter & a [K$^{-1}$] & b [K$^{-2}$] & $\frac{a}{b}$ [K] \\ \hline
$\alpha_x$ & $(9.44 \pm 0.07) \times 10^{-6}$ & $(20.4 \pm 3.7) \times 10^{-9}$ & 463 \\
$\alpha_y$ & $(12.49 \pm 0.06) \times 10^{-6}$ & $(20.0 \pm 3.0) \times 10^{-9}$ & 625 \\
$\alpha_z$ & $(-4.16 \pm 0.03) \times 10^{-6}$ & $(-12.6 \pm 1.4) \times 10^{-9}$ & 330 \\ 
\hline
\end{tabular}
\end{table}
\begin{table}[htbp]
\centering
\caption{\label{tab:beta_results_two_ways}$\beta_i$ values of RTP calculated two ways. Uncertainties are carried from $\alpha$ and reflect run-to-run variations in $\alpha$.}
\begin{tabular}{c c c c}\hline
Parameter & a [K$^{-1}$] & b [K$^{-2}$] & $\frac{a}{b}$ [K] \\ \hline
$\beta_x = \gamma_{\,zx} - \alpha_z$ & $(5.32 \pm 0.03) \times 10^{-6}$ & $(21.9 \pm 1.4) \times 10^{-9}$ & 243 \\
$\beta_x = \gamma_{\,yx} - \alpha_y$ & $(5.46 \pm 0.06) \times 10^{-6}$ & $(20.8 \pm 3.0) \times 10^{-9}$ & 263 \\
$\beta_y = \gamma_{\,zy} - \alpha_z$ & $(7.02 \pm 0.03) \times 10^{-6}$ & $(31.5 \pm 1.4) \times 10^{-9}$ & 223 \\
$\beta_y = \gamma_{\,xy} - \alpha_x$ & $(7.20 \pm 0.07) \times 10^{-6}$ & $(24.5 \pm 3.7) \times 10^{-9}$ & 253 \\
$\beta_z = \gamma_{\,yz} - \alpha_y$ & $(12.3 \pm 0.06) \times 10^{-6}$ & $(52.7 \pm 3.0) \times 10^{-9}$ & 233 \\
$\beta_z = \gamma_{\,xz} - \alpha_x$ & $(12.4 \pm 0.07) \times 10^{-6}$ & $(51.9 \pm 3.7) \times 10^{-9}$ & 239 \\
\hline
\end{tabular}
\end{table}
Using these values for $\alpha_i$ we can calculate each $\beta_j$ in two different ways, as shown in Table \ref{tab:beta_results_two_ways}. The two values are generally within the range expected from the uncertainty in $\alpha_i$ values, and the $(a/b)$ ratio is reasonably consistent, not only for the $\beta_i$ pairs, but across all three $\beta_i$s.

\section{Comparison with previously reported values for RTP}

\begin{table}[htbp]
\centering
\caption{\label{tab.rtp_comparisions} Comparison of reported values for RTP at 632.8 nm and (300 K$<T<$350 K). The first value is $(a\times 10^6)$ at $T=305$ K and the second value in parentheses is $(b\times 10^9)$. The listed $\beta_j$ are the average of the two $\beta_j$s in Table \ref{tab:beta_results_two_ways}.}
\begin{tabular}{c c c c c c}\hline
               & This            &  (i)           &  (ii)  &  (iii)     & (iv)    \\ \hline
$\alpha_x$     & 9.44(20.4)      &  10.4(8.7)     &  10.8  & 12.8(20.6) &         \\
$\alpha_y$     & 12.49(20.0)     & 14.1(11.3)     &  13.3  &            &         \\
$\alpha_z$     & $-4.16(-12.6)$  & $-4.61(-13.6)$ &$-5.9$  &            &         \\
$\beta_x$      & 5.39(21.4)      & 4.59(27.1)     &        &            & 2.90    \\
$\beta_y$      & 7.11(28.0)      & 7.08(38.1)     &        & 3.40(34.7) & 5.60    \\
$\beta_z$      & 12.35(52.3)     & 11.67(74.9)    &        & 8.74(64.3) & 10.57   \\
$\gamma_{\,xy}$  & 16.64(44.9)  & 17.48(46.8)     &        & 16.20(55.3)&         \\
$\gamma_{\,yz}$  & 24.78(76.4)  & 25.77(86.2)     &        &            &         \\
$\gamma_{\,zx}$  & 1.16(9.36)   &$-0.02(13.5)$    &        &            &         \\
$\gamma_{\,xz}$  & 21.85(72.3)  & 22.07(83.6)     &        & 21.5(84.9) &         \\
$\gamma_{\,yx}$  & 17.96(40.8)  & 18.69(38.4)     &        &            &         \\
$\gamma_{\,zy}$  & 2.86(19.0)   & 2.47(24.5)      &        &            &         \\
\hline
\end{tabular}
\begin{tabular}{l l}
(i)Mangin {\it et al}\cite{Mangin1}&
(ii)Chu {\it et al}\cite{Chu} for (293 K$<T<$373 K)\\
(iii)Yutsis {\it et al}\cite{Yutsis}&
(iv)Mikami {\it et al}\cite{Mikami}
\end{tabular}
\end{table}

Mangin {\it et al}\cite{Mangin1} measured $\alpha_x$, $\alpha_y$, and $\alpha_z$ in the range (240 K$<T<$400 K) using an optical dilatometer. They also measured all six $\gamma_{\,ij}$ values and derived $\beta$s. They did not report their $\gamma$ values, but instead gave $\alpha$ and $\beta$ values which we have used to reconstruct their six $\gamma_{ij}$. Their $b$ coefficients for $\beta$ are all larger than ours because their $b$ coefficients for $\alpha_i$ are all smaller than ours. Their $\gamma_{\,ij}$ coefficients all agree reasonably well with ours. We expect the $\gamma_{\,ij}$ measurements should agree best because they are the easiest to measure. The main disagreement arises from the $\alpha$ measurements. 

Chu {\it et al}\cite{Chu} reported values for $\alpha_x$, $\alpha_y$, and $\alpha_z$ averaged over $(293-373)$ K. If we assume that our expansion for $\alpha$ can be extended to this range, our averaged values for $\alpha$ would be ($10.0, 13.0, -4.51$)/K compared with the ($0.8, 13.3, -5.9$)/K of Chu {\it et al}.

Yutsis {\it et al}\cite{Yutsis} measured $\alpha_x$ over (273-473) K using an optical dilatometer. They also measured $\gamma_{\,xy}$ and $\gamma_{\,xz}$ over this range and derived $\beta_y$ and $\beta_z$. Their values of $\gamma$ are in close agreement with ours, but their $\alpha_x$ is substantially larger, making their $\beta$ values substantially smaller than ours. 

Mikami {\it et al}\cite{Mikami} measured $\beta_x$, $\beta_y$ and $\beta_z$ from 293-393 K every 20 K, using a prism method. They did not account for changes in apex angles due to anisotropic thermal expansion. Their $\beta$ values are substantially smaller than ours.

\section{KTP results}

Table \ref{tab:ktp_gamma_results} summarizes our KTP measurements of $\gamma_{\,ij}$ and $\gamma_i^{\;a}$. The $(\gamma_{xy}+\gamma_{yz}+\gamma_{zx}=\gamma_{yx}+\gamma_{zy}+\gamma_{xz})$ equality is verified to within 0.6\% across the full 300-350 K range for our KTP measurements. The value of $\gamma_z^{\;a}$ was too small to be reliably measured since there was only a fraction of an interference fringe over the full temperature range. Its value is left blank in the Table. We subtract the $\beta^a$ from the $\gamma_{i}^{\;a}$ of Table \ref{tab:ktp_gamma_results} to find the $\alpha_i$ values summarized in Table \ref{tab:alpha_results_ktp}.

\begin{table}[htbp]
\centering
\caption{\label{tab:ktp_gamma_results}Best-fit values of $\gamma_{\,ij}$ and $\gamma_i^{\;a}$ for KTP at 632.8 nm over 300-350 K. Here $i$ refers to the propagation direction and $j$ refers to the polarization direction. We assume $\gamma(T) = a + b\times (T - 305)$ where $T$ is in Kelvin. $\beta^a$ is our calculated thermo-optic coefficient for air at our laboratory elevation of 1600 m. We use $[n_x,n_y,n_z]=[1.7641,1.7730,1.8637]$ in deriving $\gamma_{\,ij}$.} \begin{tabular}{c c c c}\hline
Parameter & a [K$^{-1}$] & b [K$^{-2}$] & $\frac{a}{b}$[K]\\ \hline
$\gamma_{\,xy}$ & $13.12 \times 10^{-6}$ & $32.82 \times 10^{-9}$ & 400\\
$\gamma_{\,yz}$ & $18.99 \times 10^{-6}$ & $51.70 \times 10^{-9}$ & 367\\
$\gamma_{\,zx}$ & $3.68 \times 10^{-6}$ & $18.19 \times 10^{-9}$ & 202\\
$\gamma_{\,zy}$ & $5.38 \times 10^{-6}$ & $19.91 \times 10^{-9}$ & 270\\
$\gamma_{\,yx}$ & $13.26 \times 10^{-6}$ & $30.12 \times 10^{-9}$ & 440\\
$\gamma_{\,xz}$ & $17.05 \times 10^{-6}$ & $50.07 \times 10^{-9}$ & 341\\
$\gamma_x^{\;a}$ & $7.18\times 10^{-6}$ & $20.13\times 10^{-9}$ & 357\\
$\gamma_y^{\;a}$ & $8.78\times 10^{-6}$ & $25.40\times 10^{-9}$ & 346\\
$\gamma_z^{\;a}$ & $ -                $ & $-                  $ & \\
$\beta^a   $  & $-0.699 \times 10^{-6}$ & $ 3.83 \times 10^{-9}$ & \\
\hline
\end{tabular}
\end{table}
\begin{table}[htbp]
\centering
\caption{\label{tab:alpha_results_ktp}Values for $\alpha_x$, $\alpha_y$ of KTP, in the form $\alpha = a + b \times (T - 305)$.}
\begin{tabular}{c c c c }\hline
Parameter & a [K$^{-1}$] & b [K$^{-2}$] & $\frac{a}{b}$ [K] \\ \hline
$\alpha_x$ & $(7.88 \pm 0.04) \times 10^{-6}$ & $(16.3 \pm 1.7) \times 10^{-9}$ & 483 \\
$\alpha_y$ & $(9.48\pm 0.05) \times 10^{-6}$ & $(21.6\pm 2.8) \times 10^{-9}$ & 439 \\
$\alpha_z$ & $-                             $ & $-                          $ & $-$ \\
\hline
\end{tabular}
\end{table}
\begin{table}[htbp]
\centering
\caption{\label{tab:beta_results_two_ways_ktp}$\beta$s for KTP calculated two ways where possible. Uncertainties are carried from $\alpha$ and reflect the run-to-run variations in $\alpha$.}
\begin{tabular}{c c c c}\hline
Parameter & a [K$^{-1}$] & b [K$^{-2}$] & $\frac{a}{b}$ [K] \\ \hline
$\beta_x = \gamma_{\,zx} - \alpha_z$ & $- $ & $-$ & $-$ \\
$\beta_x = \gamma_{\,yx} - \alpha_y$ & $(3.78 \pm 0.05) \times 10^{-6}$ & $(8.55 \pm 2.8) \times 10^{-9}$ & 442 \\
$\beta_y = \gamma_{\,zy} - \alpha_z$ & $-$ & $-$ & $-$ \\
$\beta_y = \gamma_{\,xy} - \alpha_x$ & $(5.24 \pm 0.04) \times 10^{-6}$ & $(16.5 \pm 1.7) \times 10^{-9}$ & 316 \\
$\beta_z = \gamma_{\,yz} - \alpha_y$ & $(9.51 \pm 0.05) \times 10^{-6}$ & $(30.1 \pm 2.8) \times 10^{-9}$ & 316 \\
$\beta_z = \gamma_{\,xz} - \alpha_x$ & $(9.17 \pm 0.04) \times 10^{-6}$ & $(33.8 \pm 1.7) \times 10^{-9}$ & 271 \\
\hline
\end{tabular}
\end{table}
\begin{table}[htbp]
\centering
\caption{\label{tab:alpha_results2} Indirectly measured values for $\alpha_z$ of KTP in the form $\alpha = a + b \times (T - 305)$. The $\gamma$ values are from Table \ref{tab:ktp_gamma_results} and the $\beta$ values are from Table \ref{tab:beta_results_two_ways_ktp}.}
\begin{tabular}{c c c c }\hline
Parameter & a [K$^{-1}$] & b [K$^{-2}$] & $\frac{a}{b}$ [K] \\ \hline
$\alpha_z=(\gamma_{zy}-\beta_y)$ & $(0.14 \pm 0.04) \times 10^{-6}$ & $(3.41 \pm 1.7) \times 10^{-9}$ & 41 \\ 
$\alpha_z=(\gamma_{zx}-\beta_x)$ & $(-0.10 \pm 0.05) \times 10^{-6}$ & $(9.64 \pm 2.8) \times 10^{-9}$ & $-10.3$ \\ 
\hline
\end{tabular}
\end{table}

In Table \ref{tab:beta_results_two_ways_ktp} we deduce values of $\beta_j$ from ($\gamma_{ij}-\alpha_i$) in two ways wherever possible. Then in Table \ref{tab:alpha_results2} we deduce the value of the immeasurably small $\alpha_z$ in two ways, from ($\gamma_{\,zx}-\beta_x$) and from ($\gamma_{\,zy}-\beta_y$). 

\section{Comparison with previously reported values for KTP}

\begin{table}[htbp]
\centering
\caption{\label{tab.ktp_comparisons} Comparison of reported values for KTP at 632.8 nm and (300 K$<T<$350 K). The first value is $(a\times 10^6)$ at $T=305$ K and the second value in parentheses is $(b\times 10^9)$.}
\begin{tabular}{ccccccccc}\hline
               & This         &  (i)         &(ii)  &(iii) &(iv)&(v)          & (vi)& (vii) \\\hline
$\alpha_x$     & 7.88(16.3)   &  7.98(11.9)  &6.8   &      &9   &6.85(22)     & 7.06(8.8)&\\
$\alpha_y$     & 9.48(21.6)   &  9.69(4.1)   &9.6   &      &11  &             & 8.30(14.2) & \\
$\alpha_z$     & 0.02(6.5)    &$0.017(-0.2)$ &$-1.3$&      &0.6 &             &   &\\
$\beta_x$      & 3.78(8.55)   &  3.61(19.7)  &      &4.72  &    &             &   &4.83\\
$\beta_y$      & 5.24(16.5)   &  5.41(24.4)  &      &6.37  &    &$6.51(-2.77)$&   &6.44\\
$\beta_z$      & 9.34(32.0)   &  9.02(63.1)  &      &10.45 &    &10.68(34.9)  &   &10.42\\
$\gamma_{xy}$  & 13.12(32.8)  &  13.39(36.3) &      &      &    &13.36(19.2)  &   &\\
$\gamma_{yz}$  & 18.99(51.7)  &  18.71(67.2) &      &      &    &             &   &\\
$\gamma_{zx}$  & 3.68(18.2)   &  3.63(19.5)  &      &      &    &             &   &\\
$\gamma_{xz}$  & 17.05(50.1)  &  17.00(75.0) &      &      &    &17.53(56.9)  &   &\\
$\gamma_{yx}$  & 13.26(30.1)  &  13.30(23.8) &      &      &    &             &   &\\
$\gamma_{zy}$  & 5.38(19.9)   &  5.43(24.2)  &      &      &    &             &   &\\
\hline
\end{tabular}
\begin{tabular}{l l}
(i) Mangin {\it et al}\cite{Mangin2}&
(ii) Chu {\it et al}\cite{Chu} for (293 K$<T<$353 K)\\
(iii) Kato and Takaoka\cite{Kato}&
(iv) Bierlein and Vanharzeele\cite{Bierlein}\\
(v) Emanueli and Arie\cite{Emanueli}&
(vi) Pignatiello {\it et al}\cite{Pignatiello}\\
(vii) Wiechmann {\it et al}\cite{Wiechmann}
\end{tabular}
\end{table}

Mangin {\it et al}\cite{Mangin2} measured all three $\alpha$s using an optical dilatometer, and they measured all six $\gamma$ values at wavelengths of 1064.2, 632.8, 528.7, and 457.9 nm. Their $\gamma$ values for 632.8 nm and their $\alpha$ values are in good agreement with ours except for the $b$ coefficients of $\alpha$. This means their $\beta$ values also agree well with ours except for the $b$ coefficients. 


Chu {\it et al}\cite{Chu} reported $\alpha$ measurements averaged over the range (293 - 353) K. Their values are ($6.8, 9.6, -1.3$)/K while over this range our averaged values would be (8.17, 9.87, 0.14)/K.

Kato and Takaoka\cite{Kato} used a variety of published angle phase matching and quasi phase matching results from nonlinear mixing over a wide wavelength range to derive best fit $\beta$ values, and they present expressions for $\beta_i(\lambda)$. Their $\beta$ values at 632.8 nm are listed in Table \ref{tab.ktp_comparisons}. They list only the $a$ coefficients so their $\beta$ values are temperature independent. Adding or subtracting the same constant to the three $\beta_i(\lambda)$ curves would not change the fit to the phase matching data, assuming the contribution of thermal expansion to quasi phase matching is small. It is only the shapes of $\beta$ versus wavelength, combined with the differences in the three $\beta$ values at one wavelength that determine temperature tuning of phase matching. Kato and Takaoka's differences ($\beta_y-\beta_x$) and ($\beta_z-\beta_y$) are $1.65\times 10^{-6}$/K and $4.08\times 10^{-6}$/K, in fair agreement with our differences of $1.46 \times 10^{-6}$/K and $4.10\times 10^{-6}$/K at 305 K. Kato and Takaoka make no mention of the $\alpha_x$ value used in fitting the quasi phase matching data, but it contributes only a small change in the temperature tuning of quasi phase matching in KTP.

Emanueli and Arie\cite{Emanueli} measured $\alpha_x$ using an optical dilatometer, and they measured $\gamma_{\,xz}$ and $\gamma_{\,xy}$ at 532, 775, 787, 1064, 1509, 1545, and 1585 nm. From those results they derive expressions for $\beta_y(\lambda,T)$ and $\beta_z(\lambda,T)$ over the wavelength range 500-1700 nm. They do not report their values of $\gamma$ so we have reconstructed them at 632.8 nm. Agreement with our $\gamma$s is quite good. However, their $\alpha_x$ value is smaller than ours, which makes their $\beta$ values larger than ours by approximately $1.3\times 10^{-6}$/K.

Pignatiello {\it et al}\cite{Pignatiello} used a Moire fringe method to measure $\alpha_x$ and $\alpha_y$, and they report $a$ coefficients that are somewhat smaller than ours, and $b$ coefficients that are substantially smaller than ours. They claim an uncertainty of only $\pm 0.2\times 10^{-6}$/K on the $a$ coefficients and $\pm 1.6\times 10^{-9}$/K$^2$ on the $b$ coefficients.

Wiechmann {\it et al}\cite{Wiechmann} used a prism method to measure $\beta$, and they included the measured temperature dependence of the prism apex angle in their analysis. Their temperature range was 288-313 K. Their $\beta$ values are each larger than ours by approximately $1\times 10^{-6}$/K.

\section{Conclusion}

We measured the three linear thermal expansion coefficients and the three thermo-optic coefficients of RTP and KTP at 632.8 nm. The statistical, or run-to-run uncertainties of our measurements are indicated in the tables. We think the systematic uncertainties are less 3\%, and are due primarily to uncertainties in temperature measurements ($<1$ K), crystal length measurements ($<30$ $\mu$m), and unrecorded mode hops of the HeNe laser ($<0.01$ fringe). Our results should reduce the uncertainty arising from previously reported values. They are in good agreement with those of Mangin {\it et al}\cite{Mangin1,Mangin2} in most cases. Generally, $\gamma$ values from various reported measurements tend be in close agreement, but $\alpha$ measurements tend to have larger uncertainties, which account for most of the disagreement in the $\beta$ values. Because our $\alpha$ measurements were made using the same apparatus and techniques as our $\gamma_{ij}^{\;c}$ measurements, we believe they are quite reliable.

\end{document}